\documentclass[conference]{IEEEtran}
\IEEEoverridecommandlockouts
\usepackage{subcaption}
\usepackage{cite}
\usepackage{amsmath,amssymb,amsfonts}
\usepackage{algorithmic}
\usepackage{graphicx}
\usepackage{textcomp}
\usepackage{xcolor}
\usepackage{booktabs}
\usepackage{hyperref}
\usepackage{multirow}
\newcommand\liming[1]{}

\def\bert/{BERT}
\def\roberta/{RoBERTa}
\def\cbert/{Clinical BERT}
\def\detail/{DeTAiL}
\def\gemini/{Gemini3.5-Flash}
\def\lora/{LoRA}
\def\llamathree/{Llama3.2-1B-Instruct}
\def\qwen/{Qwen}
\def\qwentwofive/{Qwen2.5-Omni-7B}
\def\qwentwofivetext/{Qwen2.5-Instruct-7B}
\def\qwenthree/{Qwen3-Omni-30B}
\def\qwenthreefive/{Qwen3.5-VL-9B}
\def\facodec/{FACodec}
\def\fhsgold/{FHS gold 92}
\def\hifigan/{HiFiGAN}
\def\knnvc/{KNN-VC}
\def\linearvc/{LinearVC}
\def\triaanvc/{TriAAN-VC}
\def\uniaudio/{UniAudio}
\def\valle/{VALL-E}
\def\vevo/{VEVO}
\def\facodec/{FACodec}
\def\wavlm/{WavLM}
\def\whisper/{Whisper}
\def\demenba/{Demenba}
\def\effnet/{EfficientNet}

\usepackage{amsmath,amsfonts,bm}

\def\eqref#1{Eq.~\ref{#1}}

\def\1{\bm{1}}

\DeclareMathAlphabet{\mathsfit}{\encodingdefault}{\sfdefault}{m}{sl}
\SetMathAlphabet{\mathsfit}{bold}{\encodingdefault}{\sfdefault}{bx}{n}

\newcommand{\E}{\mathbb{E}}

\newcommand{\KL}{D_{\mathrm{KL}}}

\begin{document}

\title{Do Multimodal Large Language Models Need Reasoning to Classify Dementia from Speech?}


\author{\IEEEauthorblockN{Liming Wang$^1$, Neguine Rezaii$^2$, Bradford C. Dickerson$^2$, James Glass$^1$}
\IEEEauthorblockA{$^1$\textit{MIT CSAIL}, $^2$\textit{Massachusetts General Hospital and Harvard Medical School}\\
limingw@csail.mit.edu}
}

\maketitle

\begin{abstract}
Multimodal large language models (MLLMs) have emerged as a promising approach for improving the accuracy, transferability, and explainability of automatic dementia classification (ADC) systems from voice recordings. Yet it remains unclear whether their reasoning capabilities are beneficial for ADC, and how such capabilities should be leveraged. In this paper, we conduct a careful evaluation of reasoning MLLMs for ADC and show that naive strategies, such as relying on text-based rationales, can lead to hallucinated and inconsistent rationales for diagnosis and yield inferior ADC performance compared with LLM-free baselines. To overcome this limitation, we propose \textbf{De}mentia \textbf{T}hinker with Nonlinear \textbf{A}daptor and Re\textbf{i}nforcement \textbf{L}earning (DeTAiL), an adaptor-based framework that exploits the internal representations of reasoning MLLMs for improved dementia classification. Across two dementia datasets with distinct test formats and label granularities, DeTAiL consistently outperforms strong baselines and methods that rely on text-based rationales. Code is publicly available\footnote{https://github.com/cactuswiththoughts/DeTAiL.git}\footnote{Support for this research was provided by the MIT-MGB Seed Program
through the MIT HEALS. We also thank Hongyin Luo for helpful discussions.}.
\end{abstract}


\section{Introduction}
\begin{figure*}
    \centering
    \begin{subfigure}{0.49\textwidth}
        \includegraphics[width=0.99\textwidth]{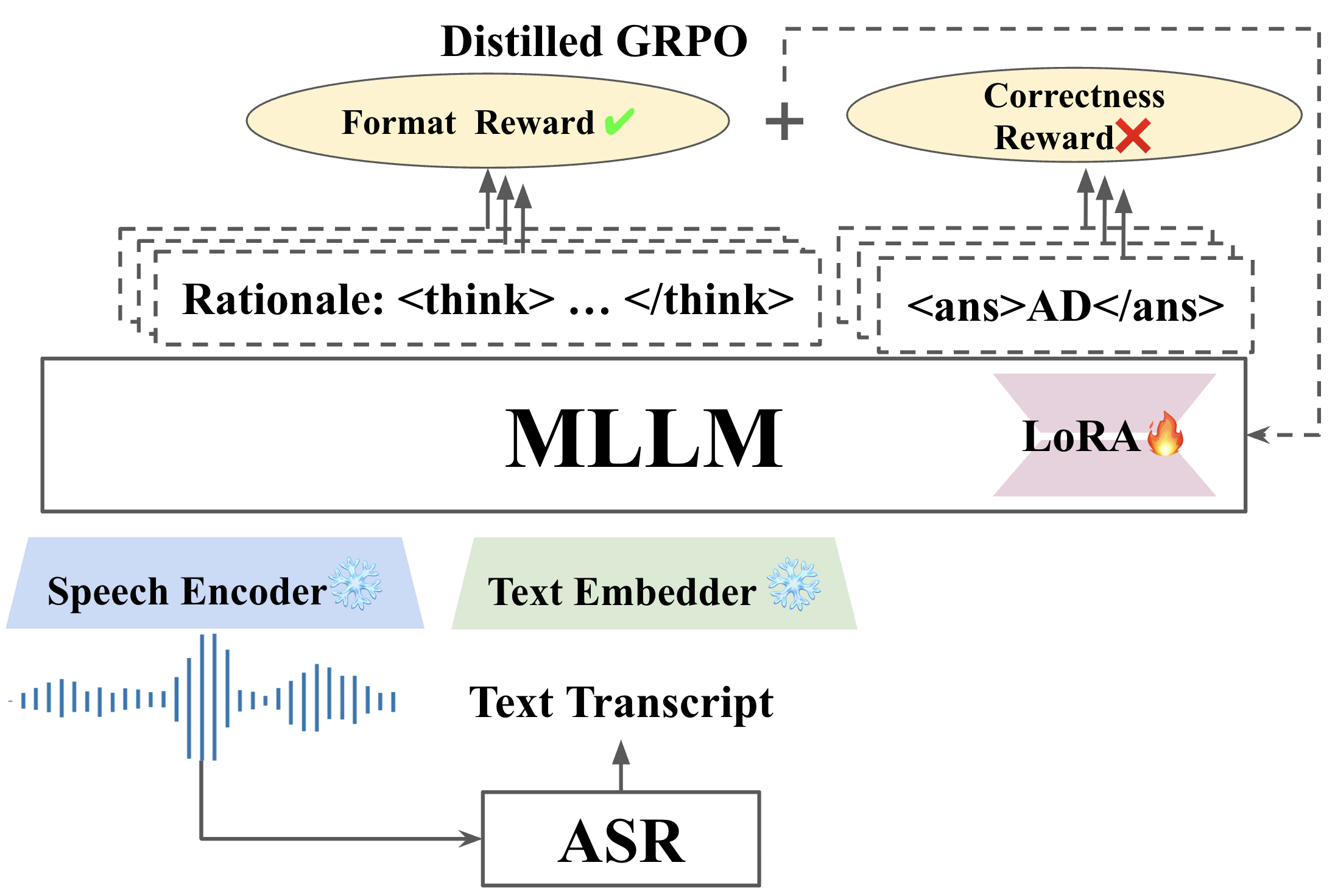}
        \caption{Distillation and GRPO stages}
    \end{subfigure}
    \begin{subfigure}{0.49\textwidth}
        \includegraphics[width=0.95\textwidth]{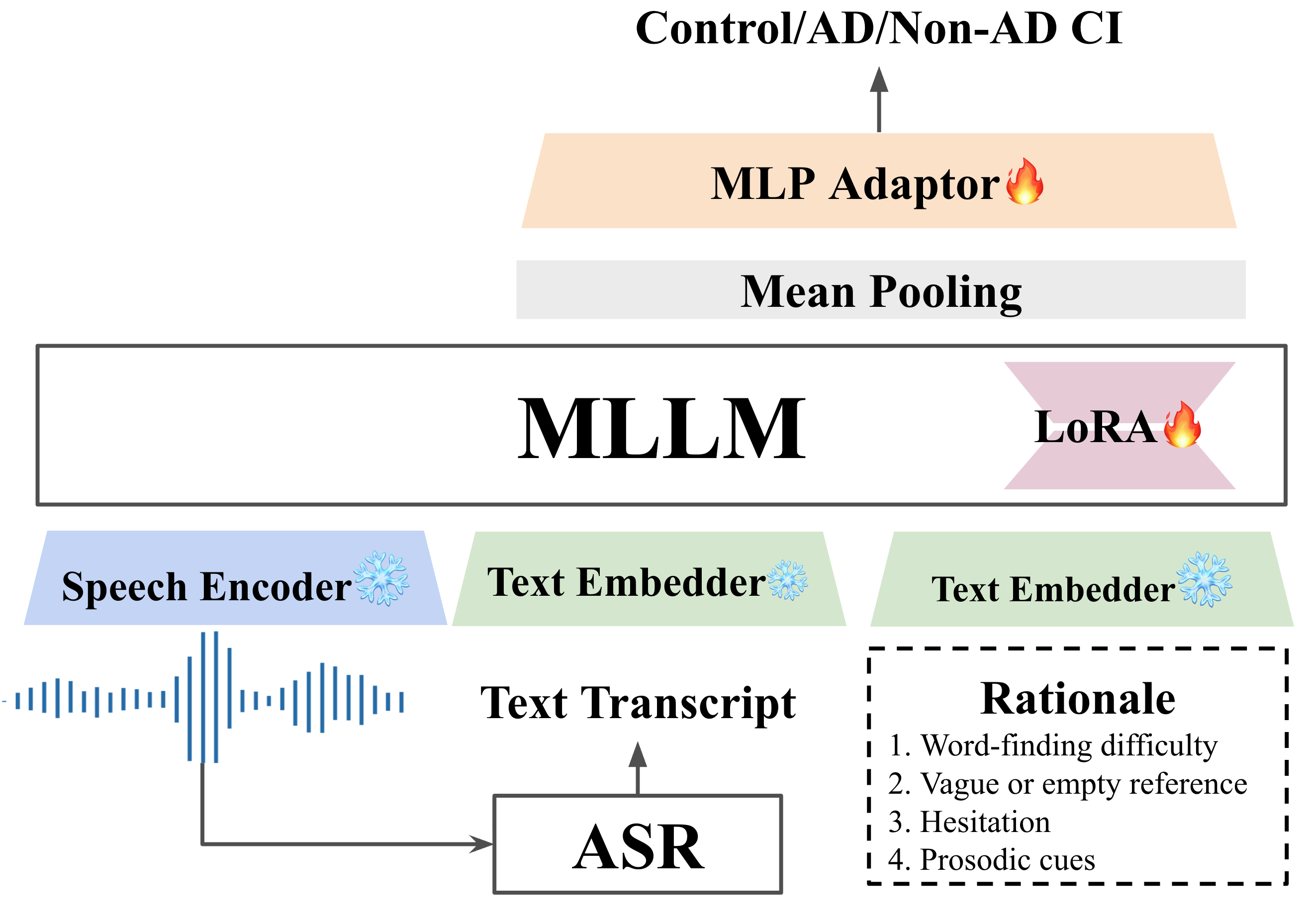}
        \caption{MLP adaptor stage}
    \end{subfigure}
    \caption{\textbf{Overall Architecture of DeTAiL.} (a) In the distillation and GRPO stages, the MLLM learns to generate both the cognitive label and the textual rationale that explains its prediction; (b) in the MLP adaptor stage, a small MLP classifier is trained on the hidden representation of the MLLM given the prompt and the generated rationale.}
    \label{fig:arch}
\end{figure*}

Speech biomarkers have emerged as an automatic, low-cost, and non-invasive option for screening and monitoring cognitive impairment and dementia, including Alzheimer's disease (AD). By listening to a patient's voice, an AI system can detect clinically useful acoustic features such as pauses, disfluencies, and prosody, as well as linguistic cues such as lexical diversity. Leveraging these features, AI-driven systems have shown promise in detecting cognitive impairment (CI) or AD~\cite{luz2020adress}, classifying disease severity~\cite{tao2025process,taherinezhad2025speechcognitive,zhang2026neurmllm}, and predicting future dementia onset~\cite{luz2021adresso}. However, most existing works on dementia classifiers often optimize accuracy, but rarely test whether and how these systems can adapt to new domains or communicate the reasoning behind their predictions.

Recent advances in large language models (LLMs) offer a promising path toward addressing these challenges.    
First, by training on large-scale free-form text data as well as text-based chain-of-thought (CoT) reasoning traces~\cite{wei2022chain}, LLMs are able to generate human-readable rationales, making it possible for them to explain their predictions using natural language. 
Further, by training on large-scale multimodal data, proprietary multimodal LLMs (MLLMs) have demonstrated strong capabilities across a range of medical domains~\cite{saab2024capabilities,achiam2023gpt4,amiri2025exploring}. 
Finally, LLMs are known for their zero-shot and few-shot capabilities, enabling them to adapt rapidly to new domains at test time~\cite{brown2020language,wei2022chain}. This creates the possibility of building general-purpose and domain-invariant screening systems for cognitive impairment and AD pathology.

Several obstacles remain, however, before this vision can be realized.
First, LLM reasoning may improve interpretability and adaptation, but may also introduce spurious rationales or distract the model from acoustic/linguistic evidence~\cite{turpin2023language,matton2025walk,afolabi2026faithful}, especially when distinguishing among fine-grained cognitive states such as dementia and mild cognitive impairment (MCI). In high-stakes settings such as medical screening, unfaithful or misleading rationales are unacceptable, as they may contribute to misdiagnosis and delay timely clinical evaluation or treatment. Prior work on LLM reasoning has explored supervised finetuning (SFT) and domain-specific post-training with reinforcement learning (RL), such as group-relative policy optimization (GRPO)~\cite{shao2024deepseekmath}. However, it remains unclear whether, and how, such reasoning-oriented post-training benefits LLM-based dementia diagnosis.
Further, MLLMs often underperform specialized models on speech classification tasks~\cite{mehta2026calm}, including dementia classification~\cite{zhang2026chainofthought}. Fully unlocking their potential for dementia diagnosis therefore remains an open challenge. This difficulty can be further exacerbated by the lack of ground-truth transcripts and by noisy speech recordings, which may introduce a domain mismatch between real-world dementia speech and the mostly clean speech data used to train many MLLMs.
In this work, we address the following research question: \emph{Do MLLMs need reasoning to classify dementia?} Our contributions are two-fold:
\begin{itemize}
    \item We train and evaluate commonly used open-source MLLMs~\cite{xu2025qwen25omni,yang2025qwen3} along three main axes:
    \begin{enumerate}
    \item \textbf{Accuracy}: We evaluate the dementia classification performance of MLLMs finetuned with various techniques, including SFT, GRPO, GRPO with rationale distillation, and a non-linear adaptor applied to the latent representations of reasoning LLMs;
    \item \textbf{Transferability}: We evaluate both zero-shot generalization and test-time adaptation (TTA) by training and testing on two dementia classification datasets with different domains and label granularities.
    \item \textbf{Explainability}: We assess whether adding rationale improves dementia classification and analyze the reliability of rationales generated by MLLMs~\cite{xu2025qwen25omni,yang2025qwen3}.
    \end{enumerate}
    \item We propose \textbf{De}mentia \textbf{T}hinker with nonlinear \textbf{A}daptor and Re\textbf{i}nforcement \textbf{L}earning (\textbf{\detail/}), an approach that better leverages information encoded in text-based rationales by probing the hidden representations of MLLMs during post-training. Across datasets, \detail/ achieves improved classification performance and transferability over standard adaptation methods such as low rank adaptors (LoRA)~\cite{hu2022lora}.
\end{itemize}
Our work reveals that while reasoning is in general beneficial for the transferability of MLLM-based dementia classifiers, whether it helps the in-domain accuracy and explainability of the model depends on its quality, modality coverage, and alignment with the target dataset.

\section{Related Work}

\textbf{Speech-based dementia classification.} 
Early work on speech-based automatic dementia classification (ADC) such as Alzheimer detection used hand-crafted speech and linguistic features~\cite{alhanai2017spoken,luz2020adress,martinc2021adress}. Other work leveraged deep learning architectures~\cite{dawalatabad-etal-2022-detecting,rohanian2021acoustic,xue2021dementia,wang2025recognizing} and pretrained speech and text foundation models~\cite{Balagopalan2021-alzheimer-w2v2,Haulcy2021-alzheimer-speech,Li2023-alzheimer-whisper}. Early work tended to focus on sentence-level binary detection (healthy or not) with datasets such as Alzheimer’s Dementia Recognition through Spontaneous Speech (ADReSS)~\cite{luz2020adress}, ADReSSo~\cite{luz2021adresso} and Framingham Heart Study (FHS)~\cite{alhanai2017spoken}, while later benchmarks such as PROCESS~\cite{tao2025process} moved toward more fine-grained classification. To enable explainability, such methods are often combined with post-hoc interpretability methods such as SHAP~\cite{lundberg2017unified,tang2023explainable}\liming{Neguine, can you provide some cites on this}.

\textbf{MLLMs for dementia and cognitive assessment.} 
 More recently, MLLMs have been explored for ADC~\cite{amiri2025exploring,casu2024optimizing,park2025reasoningcotad,zolnour2025llmcare,zhang2026neurmllm,jiang2026whatllmsknowad,wang2025recognizing, liu2026-mllm-crosslingual-ad}. Early LLM-based approaches typically prompt MLLMs directly for dementia classification using in-context learning and CoT reasoning~\cite{casu2024optimizing,amiri2025exploring,zhang2026chainofthought}, but often report performance below that of strong non-LLM baselines. Subsequent studies improved performance through SFT~\cite{liu2026-mllm-crosslingual-ad,park2025reasoningcotad} using parameter-efficient adaptation such as LoRA~\cite{hu2022lora,liu2026-mllm-crosslingual-ad} or lightweight linear adaptors~\cite{park2025reasoningcotad}. Other work has explored paired perplexity from LLMs~\cite{xiao2025alzheimers}, agentic workflows~\cite{li2025caread,kang2026agenticcognitiveprofiling,hou2026adagent}, or used MLLMs to generate synthetic data for augmenting non-LLM dementia classifiers~\cite{zolnour2025llmcare}. Beyond Alzheimer's disease, MLLMs have also been applied to other cognitive and neurodegenerative conditions, including Parkinson's disease~\cite{crawford2025parkinsonllm}, aphasia~\cite{cong2024clinical,merhbene2026detecting}, ALS~\cite{zhang2026apgrpo}, and multi-disease settings~\cite{zhang2026neurmllm,zhang2026apgrpo}. Recent studies further examine cross-lingual transfer for Alzheimer's disease classification~\cite{liu2026-mllm-crosslingual-ad} and fine-grained profiling of cognitive functions, including memory, executive function, attention, and language, to support more explainable dementia diagnoses~\cite{kang2026agenticcognitiveprofiling}.

\textbf{LLM faithfulness.}
Early work on CoT reasoning claimed that the rationale generated by the LLMs provides insights into the internal decision making process of LLMs~\cite{wei2022chain}. However, later work applied various counterfactual perturbations to the rationale and prompt and revealed the LLM explanations to be plausible yet misleading in general question answering tasks~\cite{turpin2023language,wang2023towards,matton2025walk}. \cite{afolabi2026faithful} applied similar techniques to medical reasoning LLMs and found similar trends. Others have proposed to view the LLM rationale as a form of prompt augmentation rather than a literal explanation~\cite{kambhampati2026position}. Improving LLM faithfulness remains an open challenge.

\section{Method}

In this section, we formulate the problem of \emph{explainable dementia classification} (EDC) and describe our proposed method, \textbf{\detail/}, as illustrated in Fig.~\ref{fig:arch}.
In EDC, an automatic dementia classifier (ADC) is given a multimodal prompt $q:=(x,t)$, where $x$ is the speech waveform and $t$ is either the ground-truth transcript or a transcript predicted from $x$ by an automatic speech recognizer (ASR). The goal is to predict the speaker's cognitive status $y$, which may be a coarse label such as AD or control, or a finer-grained label such as non-AD CI, AD, or control. In addition to the final predicted label $\hat{y}$, an explainable classifier generates a rationale $z$, represented as intermediate text tokens produced before the final answer.

Although a pretrained LLM can be prompted as an ADC, it often underperforms domain-specialized models. \detail/ therefore adapts a pretrained MLLM policy $\pi_{\theta}(\cdot)$ through three post-training stages: rationale distillation, RL post-training, and nonlinear hidden state adaptation.
In the \emph{distillation} stage, the model learns from a strong teacher LLM $\pi_{\theta_{\mathrm{teacher}}}$ that generates a label-conditioned teacher rationale:
\begin{align}
    z_{\mathrm{teacher}} \sim \pi_{\theta_{\mathrm{teacher}}}(\cdot \mid x,t,y),
\end{align}
where the multimodal input and gold label are wrapped using the template in Table~\ref{tab:gold-label-explanation-prompt}. The student is then trained with SFT to generate both the teacher rationale and the cognitive-status label from a prediction-time prompt that does not reveal the gold label:
\begin{align}
    \min_{\theta} -\E_{x,t,y}\log \pi_{\theta}(z_{\mathrm{teacher}},y|x,t). 
\end{align}
Instead of updating the full model, we use LoRA~\cite{hu2022lora} with rank $r=128$ and scaling factor $\alpha=32$.  
In the \emph{RL post-training} stage, we further optimize the distilled model using GRPO~\cite{shao2024deepseekmath}. Given prompt $q$, the model samples an output $o:=(\hat{z},\hat{y})$ consisting of a rationale and a label. We optimize the policy with a reward $r(o,y)$ that combines a correctness reward $r_c(\hat{y},y)$ and a format reward $r_f(o)$. The correctness reward checks whether the predicted label matches the reference label, while the format reward checks whether the output follows the ``$\texttt{<think>}\cdots\texttt{</think>}\texttt{<answer>}\cdots\texttt{</answer>}$'' structure:
\begin{align*}
    r_c(\hat{y},y)&=\begin{cases}
        1,\,\text{if }\hat{y}=y\\
        0,\,\text{otherwise}
    \end{cases},\\
    r_f(o)&=\begin{cases}
        1,\,\text{if $o$ follows the format}\\
        0,\,\text{otherwise}
    \end{cases}.
\end{align*}
For each prompt, GRPO samples a group of $G$ rollouts $o_1,\ldots,o_G$ and computes their rewards $r_1,\cdots,r_G$. This groupwise comparison provides an advantage estimate without training an explicit value function. The optimization objective is
\begin{equation}
\begin{aligned}
\min_{\theta}\ 
-\mathbb{E}_{q,o}
\frac{1}{G}
\sum_{i=1}^{G}
\frac{1}{|o_i|}
\sum_{j=1}^{|o_i|}
\left[
\ell_{i,j}^{\mathrm{clip}}(\theta)
-
\beta \KL(\pi_{\theta}\mid\mid\pi_{\theta_{\mathrm{d}}})
\right],
\end{aligned}
\end{equation}
where $\theta_d$ denotes the distilled parameters from the first stage, $\KL$ is the estimated Kullback--Leibler (KL) divergence that regularizes the updated policy against the distilled policy, $\beta$ controls the KL penalty, and
\begin{equation}
\ell_{i,j}^{\mathrm{clip}}(\theta)
:=
\min\left\{
\alpha_{i,j}(\theta)\hat A_{i,j},
\operatorname{clip}\left(\alpha_{i,j}(\theta),1-\epsilon,1+\epsilon\right)\hat A_{i,j}
\right\},
\end{equation}
for some hyperparameter $\epsilon>0$, where 
\begin{equation}
\alpha_{i,j}(\theta) :=
\frac{\pi_{\theta}(o_{i,j}\mid q,o_{i,<j})}
{\pi_{\theta_{\mathrm{d}}}(o_{i,j}\mid q,o_{i,<j})},
\end{equation}
and $\hat{A}_{i,j}\approx \frac{r_i-\mathrm{mean}(r_1,\ldots,r_G)}{\mathrm{std}(r_1,\ldots,r_G)}$ is the normalized group advantage for rollout $i$ at token position $j$. We again use LoRA in this stage.

For cross-domain evaluation, we optionally apply test-time reinforcement learning (TTRL)~\cite{zuo2025ttrl} on unlabeled target-domain samples. Specifically, we generate multiple candidate labels $\hat{y}_1,\ldots,\hat{y}_G$, obtain a pseudo-label $\tilde{y}$ by groupwise majority voting, and use $\tilde{y}$ in place of the unknown gold label when computing the correctness reward for GRPO.

\begin{table}[t]
\centering
\tiny
\caption{Prompt template for generating evidence-based explanations conditioned on a gold cognitive-status label.}
\begin{tabular}{p{0.96\linewidth}}
\hline
\textbf{Prompt template} \\
\hline
\begin{minipage}{0.94\linewidth}
\ttfamily
\textless audio\textgreater\\
You are given an audio recording, its transcript, and the gold cognitive-status label.\\[0.5em]

Your task is not to classify from scratch. Instead, explain why the provided gold label is appropriate based only on observable evidence from the audio and transcript.\\[0.5em]

Focus on evidence such as:\\
- word-finding difficulty\\
- vague or empty references\\
- semantic substitutions\\
- repetitions or perseveration\\
- disorganized narrative structure\\
- missing story elements\\
- fluency or hesitation patterns\\
- acoustic/prosodic cues, only if clearly present\\[0.5em]

Do not contradict the gold label.\\
Do not invent unsupported symptoms.\\
Do not diagnose beyond the provided label.\\
Do not mention the gold label in the explanation.\\[0.5em]

Transcript:\\
\{transcript\}\\[0.5em]

Gold label: \{gold\_label\}\\
Allowed labels: \{label\_str\}\\[0.5em]

Output exactly in this format:\\
\textless think\textgreater\\
Linguistic evidence: transcript-level observations relevant to the provided status, without naming the status.\\
Acoustic/prosodic evidence: audible speech observations; if no reliable acoustic cue is present, write ``no clear acoustic cue.''\\
Overall: one concise sentence summarizing why the observations are consistent with the provided status, without naming the status.\\
\textless /think\textgreater\\
\textless answer\textgreater\{gold\_label\}\textless /answer\textgreater
\end{minipage}
\\
\hline
\end{tabular}
\label{tab:gold-label-explanation-prompt}
\end{table}
Finally, the \emph{nonlinear adaptor} stage uses the MLLM's hidden representation rather than relying only on the generated text. We feed the post-trained MLLM either the multimodal prompt with its self-generated rationale $(q, \hat{z})$ or the multimodal prompt with the teacher rationale $(q,z_{\mathrm{teacher}})$. We then extract the last-layer hidden states $h_1,\ldots,h_L$, where $L=|q|+|\hat{z}|$, and train a small multi-layer perceptron (MLP) classifier on the mean-pooled representation:
\begin{align}
    \min_{\phi}-\E_{x,t,y}&\log p_{\phi}(y|\bar{h}),
\end{align}
where $\bar{h}=\frac{1}{L}\sum_{i=1}^L h_i$ and $\phi$ denotes the MLP parameters. This adaptor probes whether dementia-relevant information is encoded in the MLLM representation even when the textual rationale is incomplete or noisy.

\section{Experiments}
\subsection{Datasets}
We evaluate our method on two dementia speech datasets: ADReSS\cite{luz2020adress} and the  Longitudinal Early-Onset Alzheimer’s Disease Study (LEADS) dataset~\cite{rezaii2025voiceprints}. The LEADS dataset contains 188 pathology-confirmed cases and more than 300 participants in total, with confirmed cognitive and biomarker status and an average age of 57 years and an approximately equal number of males and females. We also perform a 10-fold cross validation on LEADS as it does not have a standard split. We use area under the curve (AUC) as the primary evaluation metric because it provides a threshold-independent measure of ADC performance. For LLM-based approaches without the MLP adaptor, we extract probability scores by averaging the token probabilities assigned to the answer label names: \texttt{alzheimers\_disease}, \texttt{control}, and \texttt{non\_ad\_cognitive\_impairment}. For LEADS, we additionally report AUC (CI only), which evaluates fine-grained discrimination among cognitively impaired participants, and AUC (3 class), which evaluates classification across all three classes.
For ADReSS, we use the human-annotated transcripts. For LEADS, we obtain transcripts with \whisper/-Large V3 in addition to the ground-truth transcripts. Because the ASR transcripts empirically contain more dementia-relevant information, we use them for all LEADS experiments unless otherwise specified. To obtain teacher rationales for distillation, we prompt \gemini/ with the transcript and gold label. For LEADS, we do not condition \gemini/ on raw audio in order to preserve patient privacy. As a sanity check, we train \detail/ with \qwentwofive/ using these rationales and obtain a 2-class AUC of 100\% and CI-only AUC of 93\%, suggesting that the generated rationales are largely consistent with the labels.

\subsection{Experimental Setup}
We use \qwen/-family LLM models~\cite{yang2024qwen25,xu2025qwen25omni,yang2025qwen3} for most experiments and perform SFT and GRPO with \texttt{ms-swift}~\cite{zhao2024swift}. For GRPO, we use group size $G=4$, a learning rate of \texttt{1e-5}, micro-batch size 8, two A6000 GPUs, and one training epoch. For SFT, we use a learning rate of \texttt{1e-4} and train for 20 epochs. Unless otherwise stated, all other hyperparameters follow the toolkit defaults. During inference, we use top-$p$ sampling with $p=0.99$, temperature $1$, and a maximum generation of 4096 tokens. The MLP adaptor has two fully connected layers with ReLU activation and a hidden layer size of $768$. 
As non-LLM baselines, we use ADC models based on pretrained \whisper/~\cite{Radford2023-whisper} encoder embeddings and text-language-model embeddings from \bert/~\cite{devlin2019bert} and \cbert/~\cite{alsentzer2019publicly}. For text-only baselines, we use the same MLP adaptor as in the MLLM experiments. For speech, we use \whisper/-Small as the backbone with layer-weighting and temporal pooling, following \cite{Li2023-alzheimer-whisper}. For multimodal baselines, we combine the predicted probabilities from the speech and text branches using weighted late fusion. 
\begin{table}[ht]
    \centering
    \caption{\textbf{In-domain dementia classification results on ADReSS.} For \detail/, ``teacher'' denotes conditioning on rationales generated generated by \gemini/, whereas ``self'' denotes conditioning on rationales generated by the adapted MLLM itself. Acc. denotes accuracy. Boldface and underlining denotes the best and second-best results, respectively. Models marked with ``*'' uses average result over multiple seeds for the metrics and the equal-error-rate threshold for accuracy.}
    \label{tab:adress_result}
    \resizebox{0.49\textwidth}{!}{
    \begin{tabular}{l|l|l|cc}
    \toprule
    Adaptor & Base Model & Post Training & AUC (\%) & Acc. (\%)\\
    \midrule
    \multicolumn{3}{l|}{Random Forest$^*$~\cite{martinc2021adress,xiao2025alzheimers}} & 91.3 & 82.8\\
    \midrule
    MLP~\cite{zhu2021exploring} & BERT+SpeechBERT & SFT & - & 82.9\\
    MLP & \whisper/ & SFT & 87.7 & 80.8\\
    MLP & \bert/+\whisper/ & SFT & \textbf{94.1} & \underline{87.5}\\
    \midrule
    Linear~\cite{park2025reasoningcotad} & \llamathree/ & SFT & - & \underline{87.5} \\
    None$^*$~\cite{xiao2025alzheimers} & Mistral & None & \underline{93.9} & 86.4 \\
    \midrule
    None~\cite{liu2026-mllm-crosslingual-ad} & \qwentwofive/ & CoT & - & 83.3 \\
    None & \qwentwofive/ & None & 47.6 & 52.1\\
    LoRA & \qwentwofive/ & GRPO & 86.3 & 79.0\\
    LoRA & \qwentwofive/ & Distilled SFT & 88.9 & \underline{87.5}\\
    LoRA & \qwentwofive/ & Distilled GRPO & 90.5 & 85.3\\
    \midrule
    MLP & \qwentwofive/ & None & 88.9 & \underline{87.5}\\
    MLP & \qwenthree/ & None & 89.8 & 85.4\\
    \midrule
    DeTAiL (teacher) & \qwentwofive/ & Distilled GRPO & 92.4 & 85.4\\
    DeTAiL (self) & \qwentwofive/ & Distilled GRPO & 93.6 & \textbf{89.5}\\
    \bottomrule
    \end{tabular}}
\end{table}

\begin{table*}[ht]
    \centering
    \caption{\textbf{In-domain dementia classification results on LEADS with 10-fold cross validation (CV).} We report two-class AUC, CI-only AUC for fine-grained CI discrimination, and three-class AUC. For \detail/, ``teacher'' denotes conditioning the MLLM on rationales generated by \gemini/, whereas ``self'' denotes conditioning on the rationale generated by the adapted MLLM itself. ``*'' denotes results with a different CV split from ours.}
    \label{tab:leads_result}
    \begin{tabular}{l|l|l|ccc}
    \toprule
    Adaptor & Base Model & Post Training & AUC (\%) & AUC (CI only,\%) & AUC (3 class,\%)\\
    \midrule
    \multicolumn{3}{l|}{XGBoost$^*$~\cite{rezaii2025voiceprints}} & 94.5 & 80.4 & - \\
    \midrule MLP$^*$~\cite{rezaii2025voiceprints} & \roberta/ (GT text) & SFT & 98.8 & 90.4 & - \\
    MLP & \bert/ (GT text) & SFT & 94.1$_{\pm 4.7}$ & 83.0$_{\pm 11.0}$ & 76.0$_{\pm 8.4}$\\
    MLP & \bert/ (GT text)+\whisper/ & SFT & 96.1$_{\pm 4.1}$ & 84.3$_{\pm 10.5}$ & 82.8$_{\pm 5.2}$\\
    MLP & \bert/ & SFT & 96.2$_{\pm 6.2}$ & 88.7$_{\pm 9.2}$ & 72.7$_{\pm 14.6}$ \\
    MLP & \bert/+\whisper/ & SFT & 96.3$_{\pm 6.0}$ & 89.1$_{\pm 9.2}$ & 81.9$_{\pm 9.9}$ \\
    MLP & \cbert/ & SFT & 94.5$_{\pm 4.2}$ & 90.2$_{\pm 7.4}$ & 59.0$_{\pm 5.2}$\liming{update}\\
    \midrule
    None & \qwentwofive/ & None & 77.0$_{\pm 18.0}$ & 71.9$_{\pm 13.7}$ & 71.8$_{\pm 12.2}$\\
    LoRA & \qwentwofive/ & SFT & 83.1$_{\pm 11.0}$ & 53.7$_{\pm 18.8}$ & 72.0$_{\pm 9.0}$\\
    LoRA & \qwentwofive/ & GRPO & 81.7$_{\pm 12.0}$ & 59.9$_{\pm 27.1}$ & 73.1$_{\pm 12.4}$\\
    LoRA & \qwentwofive/ & Distilled GRPO & 86.2$_{\pm 5.9}$ & 60.9$_{\pm 23.7}$ & 77.2 $_{\pm 6.8}$\\ 
    MLP & \qwentwofive/ & None & 94.4$_{\pm 6.5}$ & 83.9$_{\pm 15.5}$ & 88.7$_{\pm 7.1}$ \\
    MLP & \qwenthree/ & None & \textbf{96.6}$_{\pm 4.3}$ & \textbf{93.8}$_{\pm 7.1}$ & \textbf{91.5}$_{\pm 5.6}$\\
    \midrule
    DeTAiL (teacher) & \qwentwofive/ & Distilled GRPO & 91.9$_{\pm 7.1}$ & 84.4$_{\pm 12.4}$ & 86.1$_{\pm 6.8}$\\
    DeTAiL (self) & \qwentwofive/ & Distilled GRPO & 91.3$_{\pm 6.2}$ & 86.3$_{\pm 10.5}$ & 84.7$_{\pm 5.4}$\\
    \bottomrule
    \end{tabular}
\end{table*}
\subsection{In-domain EDC Results}
Table~\ref{tab:adress_result} reports the in-domain results on ADReSS, where we compare \detail/ with representative published baselines on ADReSS, including the best non-LLM method based on random forests~\cite{martinc2021adress} and the best LLM-based method using Mistral~\cite{xiao2025alzheimers}, though some have settings slightly different from ours and should be viewed as contextual rather than directly comparable.
The \bert/+\whisper/ baseline achieves 94.1\% AUC, improving the audio-only baseline by +6.4 absolute AUC points.
In contrast, zero-shot prompting of \qwentwofive/ performs near chance, suggesting that the pretrained MLLM does not directly expose reliable dementia-discriminative cues through label-token probabilities. This is confirmed by the fact that adding in-context learning and CoT reasoning~\cite{liu2026-mllm-crosslingual-ad} improves AUC to $83.3\%$.

Post-training substantially improves the MLLM. GRPO raises the AUC to 86.3\%, while adding rationale distillation further improves performance to 88.9\% with SFT and 90.5\% with GRPO. These results indicate that reasoning-oriented post-training can improve MLLM-based ADC, although LoRA-based adaptation still trails the strongest non-LLM multimodal baseline.

The frozen-MLLM results show a complementary pattern. Even without generative post-training, an MLP trained on MLLM hidden representations achieves 88.9\% AUC with \qwentwofive/ and 89.8\% AUC with \qwenthree/. This suggests that MLLMs encode dementia-relevant information internally even when zero-shot textual predictions are unreliable.
Finally, \detail/ further improves the MLP adaptor from 88.9\% to 93.6\% AUC and achieves the best accuracy, 89.5\%. Compared with LoRA under the same distilled-GRPO post-training, \detail/ improves AUC by 3.1 absolute points. The self-rationale variant outperforms the teacher-rationale variant on ADReSS, suggesting that rationales aligned with the base model's own generation style may provide more useful conditioning than rationales generated by a stronger external teacher.

Table~\ref{tab:leads_result} shows results on the more fine-grained LEADS task. We include published XGBoost and RoBERTa baselines~\cite{rezaii2025voiceprints} as external reference points; because they use a different cross-validation split from ours, they should be interpreted as contextual rather than directly comparable. Text-based \bert/ baselines remain strong, reaching 94.1\% AUC with ground-truth transcripts and 96.2\% AUC with ASR transcripts. Adding speech further improves the ground-truth-transcript setting from 94.1\% to 96.1\% AUC. ASR text alone also performs competitively, suggesting that the \whisper/ decoder may preserve or introduce dementia-relevant information beyond the manually transcribed text. In the three-class setting, combining ASR text with audio improves AUC from 72.7\% to 81.9\%. \cbert/ performs worse than vanilla \bert/ for the two-class task but better on the CI-only task, suggesting that clinical-domain text pretraining may be more useful for subtle distinctions among cognitively impaired groups than for separating controls from impaired participants.
For \qwentwofive/, zero-shot performance is much lower than the \bert/ baselines. SFT and GRPO improve two-class AUC into the low 80s, and distilled GRPO further improves it to 86.2\%, but LoRA-based methods remain below the non-LLM baselines. Replacing LoRA with an MLP adaptor substantially improves performance to 94.4\% AUC, indicating that the MLLM hidden states contain useful dementia representations that are not fully exploited by direct text generation. Scaling the base model further improves performance: \qwenthree/ with an MLP adaptor achieves the best results across all three LEADS metrics, including 96.6\% two-class AUC and 93.8\% CI-only AUC.

Unlike on ADReSS, adding rationale to the MLP adaptor does not uniformly improve EDC performance. \detail/ improves over the \qwentwofive/ MLP adaptor on the CI-only task, but decreases performance on two-class and three-class tasks. One possible explanation is that LEADS rationales are generated from transcripts without raw audio, and ASR noise may cause unreliable prosody-related evidence. Teacher rationales are slightly better than self-generated rationales on the two-class and three-class settings, but both remain below the best MLP-only adaptor. Overall, these results suggest that hidden-state adaptors are strong alternatives to LoRA for LLM-based EDC, while the value of explicit rationales depends on their quality, modality coverage, and alignment with the target dataset.

\begin{table}[ht]
    \centering
    \caption{\textbf{Cross-domain transferability between ADReSS and LEADS.} \qwentwofive/ is used for all LLM-based experiments.}
    \label{tab:transferability}
    \begin{tabular}{l|l|cc|c}
    \toprule
    Adaptor & Post Train. & Train set & Test set & AUC (\%) \\
    \midrule
    LoRA & GRPO & ADReSS & ADReSS & 86.3\\
    LoRA & GRPO & LEADS & LEADS & 81.7$_{\pm 12.0}$\\
    \midrule
    MLP & None & LEADS & ADReSS & 78.2$_{\pm 6.0}$ \\
    DeTAiL & Distilled GRPO & LEADS & ADReSS & \textbf{82.3}$_{\pm 5.2}$\\
    \midrule
    LoRA & GRPO & ADReSS & LEADS & 78.4$_{\pm 9.9}$\\
    LoRA & TTRL & ADReSS & LEADS & 75.6$_{\pm 12.5}$\\
    MLP & None & ADReSS & LEADS & 77.5$_{\pm 16.2}$\\
    DeTAiL & Distilled GRPO & ADReSS & LEADS & \textbf{85.3}$_{\pm 10.3}$\\
    \bottomrule
    \end{tabular}
\end{table}
\subsection{Cross-domain EDC Results}
To assess generalization under dataset shift, we train and test models across ADReSS and LEADS, as shown in Table~\ref{tab:transferability}. All models degrade when evaluated out of domain, indicating substantial mismatch in elicitation protocol, label distribution, transcript quality, or other dataset-specific factors. TTRL does not improve over the original GRPO model in the ADReSS$\rightarrow$LEADS setting, suggesting that majority-vote pseudo-labels are insufficient for reliable test-time adaptation in this task.
The MLP adaptor, despite its strong in-domain performance, transfers worse than LoRA in the cross-domain settings. This suggests that the MLP can overfit to dataset-specific hidden-state patterns. In contrast, \detail/ is more robust to domain shift, improving over the MLP by 5\% relative AUC in the LEADS$\rightarrow$ADReSS setting and by 10\% relative AUC in the ADReSS$\rightarrow$LEADS setting. These results suggest that rationale-conditioned hidden representations can improve transferability even when explicit rationales do not always improve in-domain performance.
\begin{table}[ht]
    \centering
    \caption{\textbf{Effect of input modalities using MLP adaptors.} ``A'' denotes audio, ``T'' denotes text, and ``V'' denotes vision inputs from mel-spectrogram images.}
    \begin{tabular}{l|ccc}
    \toprule
         & Dataset & AUC (\%) \\
    \midrule
    \qwentwofive/ (A) & LEADS & 66.3$_{\pm 3.9}$\\
    \qwentwofivetext/ (T) & LEADS & 87.3$_{\pm 6.4}$\\
    \qwenthreefive/ (V+T) & LEADS & \textbf{95.2}$_{\pm 6.6}$\\
    \qwentwofive/ (A+T) & LEADS & 94.4$_{\pm 6.5}$\\
    \midrule
    \qwentwofive/ (A) & ADReSS & 62.8\\
    \qwentwofivetext/ (T) & ADReSS & 84.7\\
    \qwenthreefive/ (V+T) & ADReSS & \textbf{89.6}\\
    \qwentwofive/ (A+T) & ADReSS & 88.9\\
    \bottomrule
    \end{tabular}
    \label{tab:eff_of_modalities}
\end{table}

\begin{figure}
    \centering
    \includegraphics[width=0.5\textwidth]{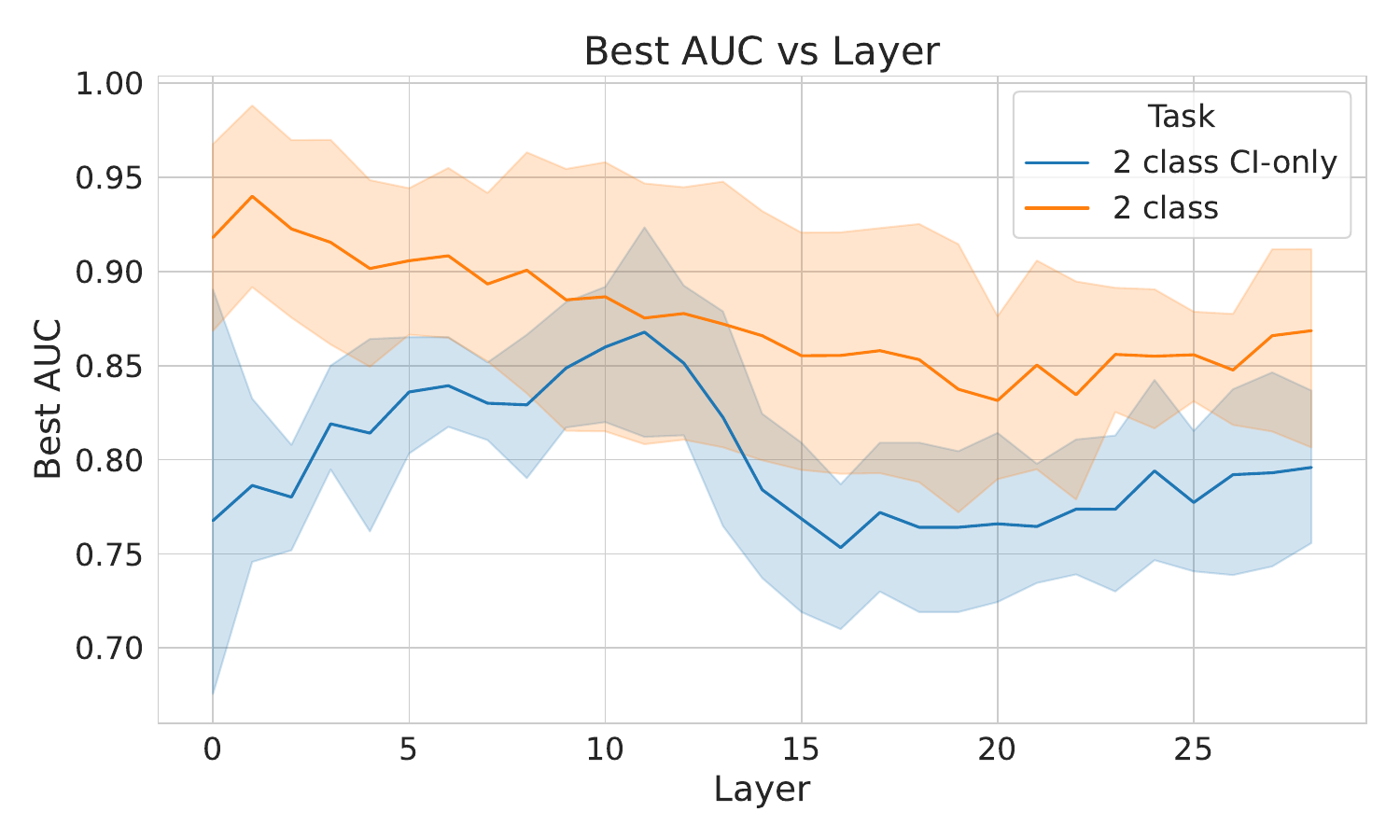}
    \caption{\textbf{Effect of MLLM layer choice for nonlinear adaptation on LEADS.} We evaluate \qwentwofive/ representations extracted from different layers with an MLP adaptor.}
    \label{fig:layerwise_analysis}
\end{figure}

\begin{figure}
    \centering
    \includegraphics[width=0.5\textwidth]{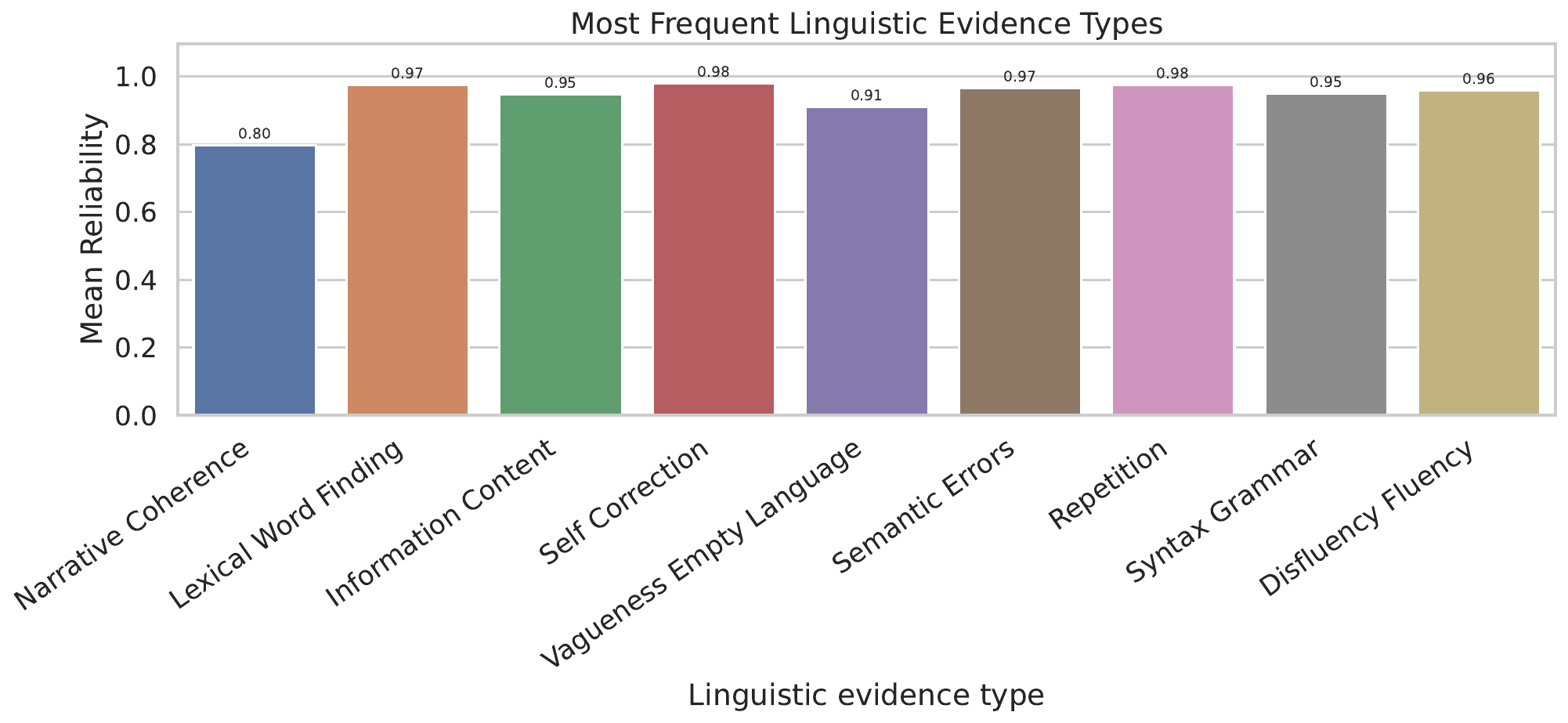}
    \caption{\textbf{Reliability of the most frequent evidence types in the rationale for \detail/ on ADReSS}. Reliability is estimated by computing the percentage of correct predictions using a given evidence.}
    \label{fig:explainability}
\end{figure}
\subsection{Ablation Analysis}
We first ablate the input modality used by the MLP adaptor in Table~\ref{tab:eff_of_modalities}. Audio-only representations perform substantially worse than text-based or multimodal representations on both datasets, whereas text and spectrogram-image inputs provide stronger signals. The vision-text setting uses the newer \qwenthreefive/ model, so its gains should be interpreted jointly as an effect of both modality and model choice. Nevertheless, the results indicate that the hidden-state adaptor can exploit dementia-relevant information across multiple input formats.

We next study which MLLM layer provides the most useful hidden representation for ADC, as shown in Fig.~\ref{fig:layerwise_analysis}. For the two-class setting, earlier layers tend to work better, suggesting that relatively low-level representations are sufficient for distinguishing controls from cognitively impaired participants. The trend differs for the CI-only setting, where the most informative representation appears around the 12th layer. Because the final layer is not consistently optimal, layer selection or layer fusion may further improve the adaptor.

Finally, we analyze the reliability of linguistic evidence types mentioned in \detail/ rationales. We classify evidence types with a keyword-matching system and estimate reliability as the percentage of correct two-class predictions among samples whose rationales mention each evidence type, as shown in Fig.~\ref{fig:explainability}. Narrative coherence is the least reliable among the frequent evidence types, whereas word repetition and self-correction are the most reliable. This pattern suggests that MLLMs rely more reliably on local speech-language patterns than on global semantic coherence for the two-class setting, consistent with the layer-wise analysis in Fig.~\ref{fig:layerwise_analysis}.

\liming{Sensitivity to hyperparameters? Effect of format reward?}
\liming{Explanability of predictions? Examples?}
\liming{Layerwise analysis for ADReSS?}
\liming{Experiment with non-qwen LLMs?}
\liming{Confusion matrix? AUROC curve?}
\liming{DeTAIL with audio only method?}
\liming{Effect of rationale teacher?}
\liming{Fusion with the non-LLM approach?}
\liming{expand and describe in text? Counterfactual analysis}

\section{Conclusion and Future Work}

We study whether reasoning helps MLLMs classify dementia from speech by evaluating its effects on accuracy, transferability, and explainability in ADC. Our results show that explicit text-based rationales are not uniformly beneficial: they can improve performance and transferability in some settings, but may also be noisy, hallucinated, or misaligned with the evidence needed for robust classification. To better leverage reasoning MLLMs, we propose \detail/, a nonlinear adaptor that uses rationale-conditioned hidden representations rather than relaying solely on generated explanations. Across two dementia speech datasets, \detail/ improves over LoRA-based adaptation and shows stronger cross-domain transfer, suggesting that MLLMs encode useful dementia-relevant information internally even when their textual rationales are imperfect.
These findings highlight both the promise and limitations of reasoning MLLMs for clinical speech analysis. Reasoning traces may help expose and transfer useful representations, but they should not be treated as faithful clinical explanations without further validation. Future work will extend \detail/ to additional domain-shift settings, including multilingual dementia speech and different elicitation protocols, and will investigate larger MLLMs, improved rationale-faithfulness evaluation, and stronger adaptation methods for reliable dementia screening support.

\bibliographystyle{IEEEtran}
\bibliography{slt_2026,llm_pathological_speech_refs,pathological_speech_multimodal_learning_refs,llm_explainability}

\end{document}